\documentclass[aps,prd,preprint,superscriptaddress,nofootinbib,showpacs,showkeys]{revtex4-2} 
\usepackage{graphicx,amsmath,amsfonts,amssymb,revsymb,dcolumn,relsize,hyperref}
\hypersetup{colorlinks=true,citecolor=blue,urlcolor=blue,linkcolor=blue}
\usepackage{mathrsfs} 

\newcommand{\be}{\begin{equation}}
\newcommand{\ee}{\end{equation}}
\newcommand{\bea}{\begin{eqnarray}}
\newcommand{\eea}{\end{eqnarray}}

\begin{document}

\title{Are  Hilbert Spaces Unphysical? Hardly, My Dear!}

\author{Nivaldo A. \surname{Lemos}} \email{nivaldolemos@id.uff.br}
\affiliation{Instituto de F\'{\i}sica, Universidade Federal Fluminense,
Av. Litor\^anea s/n \\
24210-340 Niter\'oi -- RJ, Brazil }




\renewcommand{\baselinestretch}{0.96}

\date{\today}

\begin{abstract}
It is widely accepted that the states of any quantum system are vectors in a Hilbert space. Not everyone agrees, however. The recent paper
	``The unphysicality of Hilbert spaces''  by Carcassi,  Calder\'on and Aidala is a thoughtful dissection of the mathematical structure  of quantum mechanics that seeks to pinpoint supposedly unsurmountable difficulties inherent in postulating that the physical states are elements of a Hilbert space.
Its pivotal charge against Hilbert spaces is that by a change of variables, which is a change-of-basis unitary transformation,  one ``can map states with finite
expectation values to those with infinite ones''. In the present work it is shown that this statement is incorrect and the source of the error is spotted. In consequence, the purported example of a time evolution that makes ``the expectation value oscillate from finite to infinite in finite time" is also faulty, and the assertion that Hilbert spaces ``turn a potential infinity into an actual infinity'' is unsubstantiated. Two other objections to Hilbert spaces on  physical grounds, both technically correct, are the isomorphism of separable Hilbert spaces and the unavoidable existence of infinite-expectation-value states. The former turns out to be quite irrrelevant  but the latter remains an issue without a fully satisfactory solution, although the evidence so far is that it is physically innocuous. All in all, while the authors' thesis that Hilbert spaces must be given up deserves some attention, it is a long way from being persuasive as it is  founded chiefly on a misconception and, subsidiarily, on immaterial or  flimsy arguments.

\end{abstract}




\maketitle
\newpage

\section{Introduction}

According to the first of the generally acknowledged axioms of quantum mechanics, the state of a  system at any instant is a normalized vector in a Hilbert space. There are dissenters, though. The article \cite{Carcassi} is a thought-provoking  attempt to challenge this postulate and make the case that  Hilbert spaces are not suitable to represent quantum states. Some of its criticism is not altogether without merit. For the most part, however, it falls short of its goal.

It is the purpose of the current study to  critically  address the arguments against Hilbert spaces put forward in \cite{Carcassi}. The most important objection to Hilbert space is shown to arise from  a misconception, making  it  invalid. Other secondary criticisms of Hilbert space  are  either inconsequential or rest on a fragile foundation. 

\section{The main objection to Hilbert space doesn't hold water}

First let us  focus on the central physical objection to Hilbert spaces raised by the authors of \cite{Carcassi}.  As argued below, their main line of reasoning developed  with the intention to convey that Hilbert spaces are unphysical is shattered by a fateful  misinterpretation. 

\subsection{A misinterpreted  change of variables} 

A damning physical shortcoming of Hilbert spaces, the authors  claim, is that  ``unitary transformations
in a Hilbert space can map states with finite
expectation values to those with infinite ones''. As a purported example, the change of variables $x \to y$ defined by 
\begin{equation}
\label{change-variables}
y(x) = \tan \Bigl(\frac{\pi}{2}\mbox{erf} (x) \Bigr), \qquad \mbox{erf}(x) = \frac{2}{\sqrt{\pi}}\int_0^x e^{-t^2}dt
\end{equation}
is used to transform the position probability density 
\begin{equation}
\label{prob-density-psi}
\rho_{\psi} (x)  = \vert \psi (x)\vert^2 = \frac{e^{-x^2}}{\sqrt{\pi}} 
\end{equation}
into 
\begin{equation}
\label{prob-density-phi}
\rho_{\phi} (y)  = \vert \phi (y)\vert^2 = \frac{1}{\pi (1+y^2)}. 
\end{equation}
Thus, assert the authors, through a change of variables, which is a unitary transformation representing a change of basis, the state $\psi$ in which the expectation value of any positive even power $X^n$ of the position operator is finite  has been turned into a state $\phi$ in which
these expectation values are infinite.

Clearly this cannot be true because unitary transformations corresponding to a {\it  change of basis} or {\it change of representation} in Hilbert space preserve matrix elements of any operator, which is imperative in order that physical predictions be independent of the chosen basis. This is a fundamental feature of quantum mechanics \cite{Messiah}. To make things clear, every {\it time-independent} unitary transformation in the Hilbert space of states  is a change of basis. This is because a unitary transformation maps one orthonormal basis to another orthonormal basis, and, conversely, any two orthonormal bases are connected by a unitary transformation \cite{Prugovecki}. Furthermore, any change of basis or change of representation can always be regarded as an {\it active} unitary transformation in Hilbert space that changes vectors {\it and} operators \cite{Messiah}. According to Eq. \eqref{unitary-transformation}  below, a change of variables such as \eqref{change-variables} induces a time-independent unitary transformation in the Hilbert space $L^2(\mathbb{R})$, so it is a change of basis. Now, under a change of basis carried out by the unitary operator $U$ the state vectors are transformed as $\psi^{\prime}=U\psi$ whereas the observables are transformed as ${\cal O}^{\prime}= U {\cal O}U^{\dagger}$, with 
$UU^{\dagger} = U^{\dagger}U = I$. As a result,
\begin{equation}
\label{invariance-matrix-element}
\langle\psi_1^{\prime}, {\cal O}^{\prime}\psi_2^{\prime}\rangle  = \langle U\psi_1, U{\cal O}U^{\dagger}U\psi_2\rangle = \langle \psi_1, U^{\dagger}U{\cal O}I\psi_2\rangle = \langle \psi_1, I{\cal O}\psi_2\rangle = \langle \psi_1, {\cal O}\psi_2\rangle.
\end{equation}
A fortiori, expectation values are unchanged by a change of basis performed by a unitary transformation.
Of course, things are different when it comes to the {\it unitary time evolution}, which is implemented by a {\it time-dependent} unitary operator. Then either the state vector changes while the operators remain fixed (Schr\"odinger picture)  or the state vector remains fixed while  the operators change  (Heisenberg picture), and expectation values do vary with time.

The false conclusion in \cite{Carcassi}, that a time-independent unitary transformation can make a finite expectation value infinite,   springs up from transforming the state vector but leaving the position operator untouched. Explicitly, from \eqref{change-variables} one has
\begin{equation}
\label{change-variables-derivative}
y^{\prime}(x) = \sqrt{\pi} (1+y^2) e^{-x^2}
\end{equation}
and one readily  finds that
\begin{equation}
\label{expectation-value-transformation}
\int_{-\infty}^{\infty}x^n\, \frac{e^{-x^2}}{\sqrt{\pi}}dx = \int_{-\infty}^{\infty}x^n \, \frac{e^{-x^2}}{\sqrt{\pi}}\frac{dy}{y^{\prime}(x)} = \int_{-\infty}^{\infty}\frac{x(y)^n}{\pi (1+y^2)} dy \neq \int_{-\infty}^{\infty}\frac{y^n}{\pi (1+y^2)} dy
\end{equation}
where $x(y) = \mbox{erf}^{-1}\bigl(\frac{2}{\pi}\arctan y\bigr)$,  the very same function that will be shortly shown to play the role of the transformed position operator --- refer to Eq. \eqref{position-y-multiplication} below.
It is seen that the erroneous statement in  \cite{Carcassi} is tantamount to saying that a change of variables in an integral changes the value of the integral, which of course is not the case.  
It also becomes apparent that the  unwarranted conclusion  in  \cite{Carcassi}  stems from inappropriately  interpreting the position operator in the $y$-representation as multiplication by $y$. Let us take a closer look at this question.

Consider the change of variables $y=f(x)$, where $f: \mathbb{R} \to \mathbb{R}$ is supposed to be a diffeomorphism with $f^{\prime}>0$. State vectors and operators in the $y$-representation will be distinguished by a tilde. 
The requirement
\begin{equation}
\label{change-of-variables-probability-density}
\vert \psi(x) \vert^2 dx = \vert {\tilde \psi}(y) \vert^2 dy =  \vert {\tilde \psi}(y) \vert^2 f^{\prime} (x) dx
\end{equation}
induces the definition 
\begin{equation}
\label{unitary-transformation}
{\tilde \psi} =  U \psi, \qquad  {\tilde \psi}(y) = \frac{\psi (f^{-1}(y))}{\sqrt{f^{\prime}(f^{-1}(y))}}. 
\end{equation}
This transformation is unitary because its domain and  range are the whole  Hilbert space $L^2(\mathbb{R})$, and it  preserves the inner product by construction (henceforth all integrations are over the entire real line): 
\begin{equation}
\label{inner-product-preserved}
 \int {\tilde \phi}^*(y)  {\tilde \psi}(y) dy  = \int \frac{  \phi^*(f^{-1}(y)) \psi(f^{-1}(y))}{f^{\prime}(f^{-1}(y))} f^{\prime}(x) dx = \int  \phi^*(x)  \psi (x) dx.
\end{equation}
Thus, the change of variables $y=f(x)$ is a unitary transformation implemented by the unitary operator $U$ defined by \eqref{unitary-transformation}.

Let us try to find out how the position operator in the $y$-representation should act. In the $x$-representation, the position operator $X$ is defined by
\begin{equation}
\label{position-x}
X\psi = \phi, \quad (X\psi)(x) = \phi(x) = x\psi (x).
\end{equation}
Therefore,  the position operator $\tilde X$ in the $y$-representation must be such that
\begin{equation}
\label{position-y}
{\tilde X}{\tilde \psi} = {\tilde \phi}.
\end{equation}
From \eqref{position-x}  we have 
\begin{equation}
\label{position-x-to-y-transformation}
UX\psi = U \phi \quad \Longrightarrow \quad UXU^{\dagger}U\psi = {\tilde \phi} \quad \Longrightarrow \quad UXU^{\dagger}{\tilde \psi} = {\tilde \phi}. 
\end{equation}
According to \eqref{position-y} and \eqref{position-x-to-y-transformation}, the position operator in the $y$-representation is 
\begin{equation}
\label{position-x-y-explicit}
{\tilde X} =UXU^{\dagger},
\end{equation}
which could not be otherwise in view of the general transformation law for operators previously  mentioned. Now equations \eqref{unitary-transformation},  \eqref{position-x}  and \eqref{position-y} yield
\begin{equation}
\label{position-y-multiplication}
({\tilde X}{\tilde \psi})(y) = {\tilde \phi}(y) =  \frac{\phi (f^{-1}(y))}{\sqrt{f^{\prime}(f^{-1}(y))}} =  \frac{f^{-1}(y) \psi (f^{-1}(y))}{\sqrt{f^{\prime}(f^{-1}(y))}}  =  f^{-1}(y) {\tilde \psi}(y).
\end{equation}
As it turns out, the position operator in the $y$-representation is not the multiplication operator by $y$, but the multiplication operator by $f^{-1}(y)$. Letting $Y={\tilde X}$ for a more intuitive notation, one has
\begin{equation}
\label{expectation-Y}
\langle Y \rangle  =  \int  f^{-1}(y) \vert {\tilde \psi}(y)\vert^2 dy =  \int x 
 \frac{\vert \psi (f^{-1}(y)) \vert^2}{f^{\prime}(f^{-1}(y))}  f^{\prime}(x) dx  
 =  \int x 
 \vert \psi (x) \vert^2 dx = \langle X \rangle,
\end{equation}
as it was supposed to be.

However, if the position operator in the $y$-representation is improperly taken to be multiplication by $y$, it follows that 
\begin{equation}
\label{y-expectation}
\hspace*{-0.15cm}\langle y \rangle  =  \int y \vert {\tilde \psi} (y)\vert^2 dy 
 =  \int f(x)  \frac{\vert \psi (f^{-1}(y)) \vert^2}{f^{\prime}(f^{-1}(y))}  f^{\prime}(x) dx  = \int f(x) \vert \psi (x)\vert^2 dx = \langle f(x) \rangle .
\end{equation}
Therefore, it is in fact the case that $\langle y \rangle \neq \langle x \rangle$ just because in general  $\langle x \rangle \neq \langle f(x) \rangle$. What has actually been shown in \cite{Carcassi} is that from a finite $\langle x \rangle$ one can get an infinite  $\langle y \rangle $  by picking a function $f$ that  that grows fast enough near infinity. This is correct but is of no consequence.  

There is another conceivable  way to understand the origin of the unsound conclusion in \cite{Carcassi}. The authors have interpreted the change of variables \eqref{change-variables} as a unitary transformation that changes the state vectors but leaves the observables alone, as if it were the result of a time evolution in the Schr\"odinger picture.  For  this point of view to be legitimate, the authors  would have to prove the existence of  a physically reasonable Hamiltonian such that the associated time-evolution operator achieves  the desired change of variables {\it in finite time}. Even then it would not be enough, for they would also have to perform the  hopeless task of proving  that the {\it same} Hamiltonian operator does the job for {\it all} changes of variables.  Since  they did not cogitate doing anything of the sort, their interpretation is untenable.

It is impossible to emphasize enough that a  coordinate transformation cannot change the physical predictions of any theory. This is a fundamental principle, not a matter of interpretation. One cannot simply choose to ``interpret'' a change of variables as a unitary transformation that changes the state vectors but does not change the observables. If a mere change of variables changes matrix elements and expectation values of observables then all hell breaks loose, and anything goes: quantum mechanics is reduced to a nonsensical theory devoid of any predictive power.

\subsection{A misinterpreted time-dependent change of variables} 

From the above discusssion it follows that the authors' supposed example --- equation (4) in \cite{Carcassi} --- of a time evolution that makes ``the expectation value oscillate from finite to infinite in finite time" is also flawed. This is so because, first, their concocted time-dependent coordinate transformation
\begin{equation}
\label{time-dependent-transformation}
z(x,t) = x \cos \omega t + f(x) \sin \omega t, \qquad f(x) =  \tan \Bigl(\frac{\pi}{2}\mbox{erf} (x) \Bigr)
\end{equation}
is groundlessly proclaimed to be a time evolution. And, second, this  time-dependent change of variables,  unduly portrayed  as time evolution, is misguidedly claimed to stretch ``back and forth the
wave function'' in  such a way that the position probability density  ``keeps oscillating between'' $\rho_{\psi}$  in Eq. \eqref{prob-density-psi} and $\rho_{\phi}$ in Eq. \eqref{prob-density-phi}, thereby
``making the expectation
value oscillate from finite to infinite in finite time.'' There is no doubt, therefore, that their example is based on the wrong idea that  a finite expectation value can be made infinite by a change of variables. The ensuant allegation that Hilbert spaces ``turn a potential infinity into an actual infinity'' is hence unfounded. The authors' totally unjustified and actually meaningless conflation of time evolution and time-dependent coordinate transformation is presently scrutinized in more detail.

In the Schr\"odinger picture observables are fixed in time. Thus, the only conceptually sensible interpretation of Eq. \eqref{time-dependent-transformation}  as a time evolution is that it represents the dynamical evolution of the position operator in the Heisenberg picture, namely 
\begin{equation}
\label{X-evolution-Heisenberg}
X_H (t) = X \cos \omega t + f(X) \sin \omega t,
\end{equation}
where $X$ is the position operator in the Schr\"odinger picture (the two pictures coincide at $t=0$). It just happens, however,  that the supposed time evolution \eqref{X-evolution-Heisenberg} cannot be a time evolution in the Heisenberg picture generated by 
any physically reasonable one-dimensional Hamiltonian, which takes the standard  form 
\begin{equation}
\label{standard-Hamiltonian}
H_H = \frac{P_H^2}{2} + V(X_H).
\end{equation}
Indeed, since \eqref{X-evolution-Heisenberg} implies $ X_H(0) = X$ and $\displaystyle {\dot X_H}(0) = \omega f(X)$, it follows that $X_H(0)$ and
$\displaystyle {\dot X_H}(0)$ commute. Now, the Heisenberg equation of motion for $X_H$  gives 
\begin{equation}
\label{Heisenberg-XH}
 {\dot X_H} = \frac{1}{i\hbar}[X_H,H_H] = P_H,
\end{equation}
that is, ${\dot X_H}(t) = P_H(t)$. As a consequence,
\begin{equation}
\label{commutator-contradiction}
[X_H(0),{\dot X_H}(0)] = [X_H(0),P_H(0)] = i \hbar
\end{equation}
which does not vanish. This is a contradiction.

The authors of \cite{Carcassi} seem convinced that by means of a time evolution it is possible to make a finite expectation value infinite. 
They have every right in the world to think so, but  a legitimate example would have to exhibit a time evolution operator $U(t)$ such that  the expectation value of some unbounded self-adjoint operator $\cal O$ is finite in the initial state $\psi (0)$ but is infinite in the final state $\psi (t)=U(t)\psi (0)$ for some {\it finite} time $t$.  Even if such a perverse quantum dynamics can be constructed, it will not deal a serious blow to Hilbert space  unless it is shown to arise in a case of physical interest rather than  in some artificially contrived setting.  

For what it's worth, such anomalies that are only surmised in quantum mechanics are actually  present in classical mechanics, as witnessed by the following one-dimensional  examples. (A) Let the force on a unit-mass particle be $F={\dot x}^2$ and consider the motion   with initial conditions $x(0)=0,\, {\dot x}(0)=1$. Then the solution to Newton's equation of motion ${\ddot x}={\dot x}^2$ is $x(t) = - \ln (1-t)$, and the position becomes infinite in the finite time $t=1$.  (B) Let the force be $F=6x^{1/3}$ and suppose the  initial conditions are $x(0)=0, \, {\dot x}(0)=0$. One obvious solution to Newton's equation of motion ${\ddot x}=6x^{1/3}$ is $x(t) \equiv 0$. Contrary to expectation,  $x(t)=t^3$ is another solution, and uniqueness is lost. 

In Newtonian mechanics  these pathologies are expected to be  absent in  physically realistic situations. The wish expressed in \cite{Carcassi} that the   ``mathematical spaces used in physics should already come
equipped with the proper structure that excludes physically
pathological behavior'' seems hard to be fulfilled even in Newtonian mechanics.

\section{Isomorphism of separable Hilbert spaces}

It is  somewhat surprising  that all separable Hilbert spaces are isomomorphic, meaning that any two of them are related by a unitary transformation. This implies that $L^2(\mathbb{R}^{3n})$, the state space for an $n$-particle system,  appears to be mathematically the same as $L^2(\mathbb{R})$, the state space for a single particle in one dimension. This is mentioned in \cite{Carcassi} as an undesirable and unphysical property of Hilbert spaces.  Indeed, at first sight this isomorphism of separable Hilbert spaces looks weird from the physical point of view. Yet on a closer scrutiny this impression swiftly dissipates. A particular realization of a separable Hilbert space may enjoy special features, making room for certain additional  structures that do not have a natural counterpart in other realizations. Furthermore, a quantum system is not defined solely by the Hilbert space  but also by the observables, particularly  the Hamiltonian operator, which generates the dynamics. By way of illustration,  for the hydrogen atom  with the proton fixed  the Hilbert space is $L^2(\mathbb{R}^3)$ with Hamiltonian operator $H_{\scriptstyle  hyd}$, whereas for the harmonic oscillator the Hilbert space is $L^2(\mathbb{R})$ with Hamiltonian operator $H_{\scriptstyle  osc}$. Now, unitarily equivalent operators have the same spectrum. This implies that
$H_{\scriptstyle  hyd} $ and  $H_{\scriptstyle osc}$  are not unitarily equivalent because their spectra are different. Hence, the hydrogen atom  and the harmonic oscillator are physically distinct quantum systems regardless of the fact that $L^2(\mathbb{R}^3)$ and  $L^2(\mathbb{R})$ are isomorphic.

Even in quantum field theory, which handles  systems with infinitely many degrees of freedom, the state space is postulated to be a separable Hilbert space \cite{Streater}. Surely this does not mean that, as far as the physical content is concerned, the standard model of elementary particle physics can be encoded into the particle in a one-dimensional box. 
Therefore, the contention that the isomorphism of  separable Hilbert spaces detracts from their physicality  carries no weight.

\section{Infinite expectation values}

Since most observables in quantum mechanics are represented by unbounded self-adjoint operators, whose domain  cannot be the whole Hilbert space, states characterized by infinite expectation values are unavoidable. This is disapprovingly highlighted in \cite{Carcassi}, and may be looked upon as a  disquieting aspect to quantum mechanics in Hilbert space. Just like in Newtonian mechanics, the conventional way out runs as follows: for any prepared or naturally occurring physical system, on which  measurements can be performed, it is taken for granted that infinite expectation values do not come about, that is, infinite-expectation-value elements of Hilbert space do not show up as physical states of the system. Admittedly, this is not the most desirable state of affairs. Ideally, the mathematical description of a physical theory should ``not allow unphysical objects", only those ``physical entities we prepare in a lab'' should be mathematically represented \cite{Carcassi}. 
It appears, though, that it is hardly possible to prevent the mathematical formalism of a  physical theory from containing  elements that cannot be physically realized. In truth, according to  Heisenberg's recollections \cite{Heisenberg},  in 1926 he was taught by Einstein that  ``it is quite wrong to try founding a theory on observable magnitudes alone.'' For example, a point electric charge and point dipoles are idealized (unphysical) elements of classical electrodynamics. The infinite self-energy  of and the radiation reaction force on a point electric charge pose a deep conceptual problem that has not so far been satisfactorily solved. In the words of  Jackson \cite{Jackson},  ``partial solutions, workable within limited areas, can be given'' but ``the basic problem remains unsolved''. In spite of this, classical electrodynamics, without departure from its standard mathematical framework,  is a highly successful theory, and there is no hint  that any of its measurable predictions regarding physically realizable devices has ever been jeopardized by this awkward conceptual blemish.  
Created as they are by imperfect human beings, it is probably too much to demand that physical theories and their mathematical underpinnings be perfect.

Imperfections notwithstanding, it should not go unmentioned that the self-adjointness of atomic and molecular  Hamiltonians, proved by Kato \cite{Kato}, enhances the confidence in the physicality of Hilbert spaces, and strongly indicates  that infinite expectation values are a mathematical nuisance without physical consequences.

 It falls to those who avow that Hilbert spaces are unphysical because of  the existence of infinite expectation values to produce a genuine example of the presumed unphysicality.

\section{The plea for Schwartz space}

In \cite{Carcassi} the Schwartz space   $\mathscr{S}(\mathbb{R}^{3n})$ of infinitely differentiable rapidly decreasing functions is endorsed as  ``a much more
reasonable candidate to capture the physics'' of an $n$-particle quantum system than the standard state space $L^2(\mathbb{R}^{3n})$. There is no doubt that all expectations of $\frac{1}{2}(X^nP^m + P^mX^n)$ on  $\mathscr{S}(\mathbb{R})$ are finite, as required in \cite{Carcassi}. Despite appearences,  Schwartz space is not a  remedy that heals  all  maladies. Although all expectation values of the formally self-adjoint operator $T = X^3P + PX^3$ on $\mathscr{S}(\mathbb{R})$ are finite, $T$ is only a symmetric operator that 
admits no self-adjoint extension \cite{Bogolubov}.

 In addition to that, the seemingly harmless idea of replacing $L^2(\mathbb{R}^{3n})$ with $\mathscr{S}(\mathbb{R}^{3n})$ shakes the mathematical foundations of nonrelativistic quantum mechanics,  and doing so opens up a Pandora's box of quandaries that cannot be ignored. The proposal  that Hilbert space be given up raises serious concerns about the fate of self-adjoint operators and the spectral theorem. This theorem states that there is a unique projection-valued measure  associated with each self-adjoint operator on a Hilbert space  \cite{Hall}, which is crucial to ensure that the physical predictions of quantum mechanics are unambiguous. On the face of it, completeness  seems indispensable for the validity of the spectral theorem.  It appears problematic that  on Schwartz space the position and momentum operators  are only essentially self-adjoint. It is hard to envisage how the notion of  essential 
self-adjointness can be defined without referring to some extension of the operator's domain, which would inevitably involve going beyond Schwartz space. This might offer  food for  thought to mathematicians. Anyway, perhaps --- and this is a big perhaps --- some  sort of intrinsic essential self-adjointness can be defined and shown to be sufficient to establish the spectral theorem  without the necessity  of resorting to the operator's unique self-adjoint extension.  If this proves to be the case, one of the basic tenets  of quantum mechanics could be  rephrased to state that   to each measurable quantity there corresponds an essentially self-adjoint operator. It remains to be seen whether this is a fruitful line of inquiry. 

Be that as it may, the lack of a generalization of Schwartz space to systems with infinitely many degrees of freedom weakens the proposal that Schwartz spaces replace  Hilbert spaces in general. It sounds odd that a separable Hilbert space is fine for quantum field theory but is no good for nonrelativistic quantum mechanics. After all, for systems of identical particles, nonrelativistic quantum mechanics can be equivalently  formulated as a quantum field theory by promoting the Schr\"odinger wave function to a field operator \cite{Schiff,Huang}.

\section{Conclusion}

To sum up, the cardinal argument for the standpoint espoused in \cite{Carcassi} that Hilbert spaces are unphysical is  belied by a  fallacy derived from a misconception. The allegation that Hilbert spaces are unphysical because all separable Hilbert spaces are  isomorphic  is beside the point. Another objection  to Hilbert spaces, namely the existence of infinite-expectation-value states, is really embarrassing and does not seem to have a completely satisfactory answer. Nonetheless,  all signs point in the direction that this mathematical defect is  physically inoffensive.  Finally, regarding the suggestion that the standard Hilbert space  $L^2(\mathbb{R}^{3n})$ be replaced with the Schwartz space $\mathscr{S}(\mathbb{R}^{3n})$ as the state space of an $n$-particle system, it does not  cure all mathematical illnesses and  besides lacks convincing motivation considering how feeble are the alleged evidences produced against the physicality of Hilbert spaces. At the time of writing, in March 2025, the   year that marks the hundredth anniversary of quantum mechanics,   no cogent reason to revise its first axiom is in sight.


\vspace{0.7cm}

\noindent {\bf \sf Author contributions} \hspace{0.1cm} N. A. L. did everything. 

\vspace{0.5cm}

\noindent  {\bf \sf Funding} \hspace{0.1cm} There was no funding.

\vspace{0.5cm}

\noindent  {\bf \sf Data availability} \hspace{0.1cm} No datasets were generated or analyzed during the current study.

\vspace{1cm}

\noindent {\large {\bf  \sf  Declarations}}

\vspace{0.5cm}

\noindent {\bf \sf Competing interests} \hspace{0.1cm} The authors declare no competing interests.

\end{document}